\documentclass[prl,aps,reprint,amsmath,amssymb,superscriptaddress]{revtex4-2}
\usepackage{graphicx}
\usepackage{physics}
\usepackage{hyperref}
\usepackage[compress]{cleveref}
\usepackage[dvipsnames]{xcolor}

\usepackage{soul}

\renewcommand{\vec}[1]{\mathbf{#1}}

\def\uu {{\mathbf{u}}}

\def\aa {{\mathbf{a}}}
\def\rr {{\mathbf{r}}}

\begin{document}

\title{Lagrangian irreversibility and energy exchanges in rotating-stratified turbulent flows}

\author{S. Gallon}
\affiliation{Univ Lyon, ENS de Lyon, CNRS, Laboratoire de Physique, F-69342 Lyon, France}
\author{A. Sozza}
\affiliation{Univ Lyon, ENS de Lyon, CNRS, Laboratoire de Physique, F-69342 Lyon, France}
\affiliation{CNR, Institute of Atmospheric Sciences and Climate, 10133 Torino, Italy}
\author{F. Feraco}
\affiliation{Univ Lyon, CNRS, École Centrale de Lyon, INSA de Lyon, Univ Claude Bernard Lyon 1, Laboratoire de Mécanique des Fluides et d’Acoustique - UMR 5509, F-69134 Écully, France}
\affiliation{Leibniz-Institute of Atmospheric Physics at the Rostock University, 18225 Kühlungsborn, Germany}
\author{R. Marino}
\affiliation{Univ Lyon, CNRS, École Centrale de Lyon, INSA de Lyon, Univ Claude Bernard Lyon 1, Laboratoire de Mécanique des Fluides et d’Acoustique - UMR 5509, F-69134 Écully, France}
\author{A. Pumir}
\affiliation{Univ Lyon, ENS de Lyon, CNRS, Laboratoire de Physique, F-69342 Lyon, France}
\affiliation{Max Planck Institute for Dynamics and Self-Organization, 37077 Göttingen, Germany}
\date{January 25, 2024}

\begin{abstract}
Turbulence in stratified and rotating turbulent flows is characterized by an interplay between waves and eddies, resulting in continuous exchanges between potential and kinetic energy. Here, we study how these processes affect the turbulent energy cascade from large to small scales, which manifests itself by an irreversible evolution of the relative kinetic energy between two tracer particles. We find that when $r_0$, the separation between particles, is below a characteristic length $\ell_t$, potential energy is on average transferred to kinetic energy, reducing time irreversibility, and conversely when $r_0 > \ell_t$. Our study reveals that the scale $\ell_t$ coincides with the buoyancy length scale $L_B$ over a broad range of configurations until a transitional wave-dominated regime is reached.
\end{abstract}
\maketitle
The dynamics of rotating and stratified flows, as they occur in the oceans on Earth, in planetary atmospheres and in stars, depend strongly on their rotation rate, natural buoyancy frequency and the ratio between these physical parameters~\cite{pedlosky82,davidsonTurbulence2013}. These flows in nature are in general highly turbulent, with estimated Reynolds numbers of at least $\mathrm{Re} > 10^9$ for terrestrial oceans~\cite{thorpeIntroduction2007}, involving a wide range of spatial scales, from the largest flow structures of the order of tens of kilometers (or more in astrophysical frameworks \cite{mieschLargeScale2005}), down to dissipative scales of a few millimeters~\cite{moninStatistical2013,frischTurbulence1995,thorpeIntroduction2007}. Much of our understanding of turbulent flows comes from studying the idealized case of three-dimensional homogeneous isotropic turbulence (HIT), which has been postulated to describe the  properties of small-scale motions in the limit of very large Reynolds numbers~\cite{taylorStatistical1935,kolmogorovLocal1941}. In this case, flows are characterized by a flux of kinetic energy from large to small scales, where dissipation occurs at a rate $\varepsilon_\nu$. The resulting cascade of energy introduces a fundamental irreversibility of the flow, which has a clear signature on the relative motion of two tracer particles that are initially separated by a distance $r_0 = | \rr(0) |$. 
This property rests on the identity relating the average of the scalar product between their velocity and acceleration differences to the energy dissipation rate $\langle \Delta_{\rr_0} \uu \cdot \Delta_{\rr_0} \aa \rangle \approx - 2 \varepsilon_\nu$, with the Lagrangian acceleration $\aa = \mathrm{D}/\mathrm{D}t \ \uu$, provided $r_0$ is in so-called inertial range~\cite{mannExperimental1999,falkovichParticles2001,pumirLagrangian2001}. This relation 
implies that the relative kinetic energy of pairs of tracers, as well as their separation $\langle (\rr(t) - \rr_0 )^2 \rangle$, are not even functions of time, which allows us to distinguish the time evolutions forwards and backwards in time~\cite{juchaTimereversalsymmetry2014,cheminetEulerian2022}. We notice that other manifestations of the time-irreversibility have been obtained from Lagrangian turbulence~\cite{xuFlight2014,juchaTimereversalsymmetry2014}.

In geophysical frameworks, turbulent flows are subject to stratification (STRAT) and rotation (ROT), which introduce waves that couple velocity (kinetic energy) and density fluctuations (potential energy)~\cite{Pouquet+19,davidsonTurbulence2013}. These waves induce flow instabilities, and lead to the formation of strong vertical drafts, and to turbulence as a result of wave breaking~\cite{roraiTurbulence2014,dauxoisInstabilities2018,feracoConnecting2021,caulfieldLayering2021,marinoTurbulence2022,taylorSubmesoscale2023}. Furthermore, they affect the inter-scale energy transfer leading in some cases to inverse or dual kinetic and/or potential energy cascades \cite{marino2013,marino2015,alexakisCascades2018}. Here, we investigate the interaction between kinetic $\uu^2/2$ and potential energy $\theta^2/2$ in the flow, their effect on the turbulent cascade and the resulting irreversibility. We revisit the energy budget for such flows
using the Boussinesq approximation~\cite{davidsonTurbulence2013}, thereby generalizing the Karman-Howarth-Monin relations  to the Lagrangian framework~\cite{sozzaDimensional2015}. We establish that the time-symmetry breaking term involves not only the energy dissipation $\varepsilon_\nu$, but also an exchange term between potential and kinetic energy. 

Numerically, we study the influence of waves and turbulence, keeping fixed the ratio between stratification and rotation in a configuration of oceanographic relevance~\cite{garabatoWidespread2004,nikurashinRoutes2013}. In a flow forced by injecting kinetic energy, we observe that at large scales, energy is transferred on average from kinetic to potential energy. We find at smaller scales an inversion of the flux, i.e. from potential to kinetic energy, which is in qualitative agreement with previous numerical observations in spectral space~\cite{hollowayBuoyancy1988,staquetStatistical1998,carnevaleBuoyancy2001,brethouwerScaling2007,roraiStably2015}. This inversion allows us to identify a characteristic length scale $\ell_t$ of the flow, which we  relate to the typical vertical length scale of stratification layers and therefore the complex interactions of waves and vortices~\cite{billantSelfsimilarity2001,brethouwerScaling2007,waiteStratified2011}.

\begin{table*}[t!] 
    \caption{Parameters of the runs. $N$ Brunt-V\"ais\"ail\"a frequency, $\nu$ kinematic viscosity, $\varepsilon_\nu$ kinetic energy dissipation rate (averaged over particle integration time), $\mathrm{Fr}$ Froude number, $\mathrm{Re}$ Reynolds number, $R_{\mathrm{IB}}$ Buoyancy Reynolds number, $\ell_O$ Ozmidov scale. Runs A0 - A4 where run at a grid resolution of $M=512$, runs B0 - B4 at $M = 1024$. We set the viscosity to $\nu = 1.5 \cdot 10^{-3}$ for run A0, $\nu = 1 \cdot 10^{-3}$ for runs A0-A4 and $\nu = 2.1 \cdot 10^{-4}$ for runs B0-B1. We integrated the trajectories of $1.5 \cdot 10^6$ particles for runs A0-A4 and $ 6 \cdot 10^6$ particles for runs B0-B4. All simulations have been run for at least $4.5$ eddy turnover times after the insertion of the particles.} \label{tab:param}
    \begin{ruledtabular}
        \begin{tabular}{lllllllllll}
Run & A0 & A1 & A2 & A3 & A4 & B0 & B1 & B2 & B3 & B4 \\
$N$ & 0 & 2.95 & 4.92 & 7.37 & 14.7 & 0 & 1.18 & 2.95 & 7.37 & 14.7 \\
$\varepsilon_\nu$ & 0.375 & 0.206 & 0.156 & 0.123 & 0.0237 & 0.00784 & 0.00932 & 0.00715 & 0.0136 & 0.0344 \\
$\mathrm{Fr}$ & $\infty$ & 0.169 & 0.105 & 0.0777 & 0.0410 & $\infty$ & 0.155 & 0.0823 & 0.0599 & 0.0356 \\
$\mathrm{Re}$ & 2380 & 3140 & 3270 & 3620 & 3820 & 4790 & 5510 & 7300 & 13300 & 15800 \\
$R_{\mathrm{IB}}$ & $\infty$ & 23.8 & 6.47 & 2.26 & 0.109 & $\infty$ & 31.9 & 3.92 & 1.19 & 0.755 \\
$\eta$ & 0.00974 & 0.00834 & 0.00894 & 0.00950 & 0.0143 & 0.00586 & 0.00561 & 0.00600 & 0.00511 & 0.00405 \\
$\ell_O$ & $\infty$ & 0.0700 & 0.0318 & 0.0167 & 0.00161 & $\infty$ & 0.00791 & 0.00243 & 0.00185 & 0.00234 \\
$L_B$ & $\infty$ & 0.424 & 0.264 & 0.195 & 0.103 & $\infty$ & 0.390 & 0.207 & 0.151 & 0.0896 \\
\end{tabular}
\end{ruledtabular}
\end{table*}

We perform direct numerical simulations (DNS) of the Boussinesq 
equations in a rotating frame, with constant solid body rotation 
rate $\Omega$ (and frequency $f=2\Omega$) and gravity $\vec{g}$ 
anti-aligned in the vertical ($z$) direction. 
The fluid is linearly stably stratified, subjected to a mean 
density profile $\overline{\rho}=\rho_0 -\gamma z$, 
parameterized by the Brunt-V\"ais\"al\"a frequency 
$N=(\gamma g/\rho_0)^{1/2}$.
The equations of motion for the incompressible velocity field 
$\vec{u}(\vec{x},t)=(u,v,w)$ (e.g. $\divergence{\vec{u}}=0$) 
and the density fluctuation field 
$\theta(\vec{x},t)=(\gamma/N)(\rho-\overline{\rho})$ read
\begin{align}
\label{eq:Bous_u}
\partial_t \vec{u}  + \vec{u} \cdot \boldsymbol{\nabla} \vec{u} &= 
-\grad p - f \vec{e}_z  \cross \vec{u}  \\
&\quad - N \theta \vec{e}_z 
+ \nu \laplacian \vec{u} + \vec{\Phi} \qq*{,} \notag \\
\label{eq:Bous_T}
\partial_t \theta + \vec{u} \cdot \grad \theta &= 
N w + \kappa \laplacian \theta \qq*{,}
\end{align}
where $p(\vec{x},t)$ denotes the reduced pressure, $\nu$ the kinematic 
viscosity and $\kappa$ thermal diffusivity. Here, we choose $\nu = \kappa$, or equivalently, $\mathrm{Pr} = \nu/\kappa = 1$. \Cref{eq:Bous_T,eq:Bous_u} are integrated in a triply periodic box of resolution $M=512,1024$ by means of the pseudo-spectral solver GHOST (Geophysical High-Order Suite for
Turbulence)~\cite{mininniHybrid2011} with a $2$-nd order explicit Runge-Kutta scheme for the time stepping. To sustain turbulence and to achieve a statistically stationary state, an external mechanical isotropic forcing $\vec{\Phi}(\vec{x},t)$ is included to inject energy at large scale in a narrow wave number band within the range $2 \le k_{\Phi} \le 3$, hence the resulting characteristic forcing length scale $L_{\Phi} = 2\pi/2.5$. The rate of kinetic energy injection reads $\varepsilon_{\Phi} = \langle \vec{u} \cdot \vec{\Phi} \rangle$.
To quantify the relative strength of turbulence, stratification and rotation, we consider the dimensionless Reynolds ($\mathrm{Re}$), Froude ($\mathrm{Fr}$) and Rossby ($\mathrm{Ro}$) numbers at forcing scale $L_{\Phi}$, given by
\begin{equation}
    \mathrm{Re} = \frac{u' L_{\Phi}}{\nu} \qq*{,} \mathrm{Fr} = \frac{u'}{L_{\Phi} N} \qq*{,} \mathrm{Ro} = \frac{u'}{L_{\Phi} f} \qq*{,}
\end{equation}
where $u'$, defined as the root mean square of the velocity fluctuation $u' = \langle \abs{\vec{u}}^2 \rangle^{1/2}$, is the characteristic velocity of the large-scale flow. The ratio $N/f = \mathrm{Ro}/\mathrm{Fr}$ measures the relative strength of stratification and rotation. In the present simulations, we chose to fix $N/f = 5$, which is found in the Southern Ocean~\cite{garabatoWidespread2004,nikurashinRoutes2013}. The typical vertical length scale of stratification layers is given by the buoyancy scale~\cite{billantSelfsimilarity2001,brethouwerScaling2007}
\begin{equation}
    L_B = u' / N \qq*{.}
\end{equation}
To compare the relative strength of stratification and dissipation on small scales, i.e. the relative importance of internal-gravity waves and vortices, we define the buoyancy Reynolds number $R_\mathrm{IB} = \varepsilon_\nu / (\nu N^2)$~\cite{davidsonTurbulence2013}, which is related to the ratio between the Ozmidov scale $\ell_O=(\varepsilon_\nu/N^3)^{1/2}$ 
and the Kolmogorov scale $\eta=(\nu^3/\varepsilon_\nu)^{1/4}$ 
as $R_\mathrm{IB} = (\ell_O/\eta)^{4/3}$. The former is the length scale below which stratification effects become negligible and the flow recovers isotropy, while the latter is the characteristic scale for dissipation. Alternatively, $R_\mathrm{IB}$ can be interpreted as a ratio between times as $R_\mathrm{IB}=(\tau_\eta N)^{-2}$, namely the viscous time $\tau_\eta=(\nu/\varepsilon_\nu)^{1/2}$ and the typical time of the internal gravity waves $\sim 1/N$. For the dynamics of single tracer particles, at transition between a wave-dominated ($R_{\mathrm{IB}}<0$) and an eddy-dominated regime ($R_{\mathrm{IB}}>0$) was recently observed~\cite{buariaSingleparticle2020}. 
The buoyancy Reynolds number $\mathrm{R}_\mathrm{IB}$ used here is related to, but differs from $R_\mathcal{B} = \mathrm{Re} \mathrm{Fr}^2$, see~\cite{iveyDensity2008}.

To ensure that our simulations are satisfactorily resolved, we kept $k_\mathrm{max}\eta \gtrsim 2$, where  $k_\mathrm{max}=M/3$ is the largest wavenumber in the simulation. 
We choose a time increment $\mathrm{d}t$ sufficiently small so that the fastest waves are well resolved.
The simulation parameters are summarized in \cref{tab:param}.

\begin{figure*}[t!]
\includegraphics[]{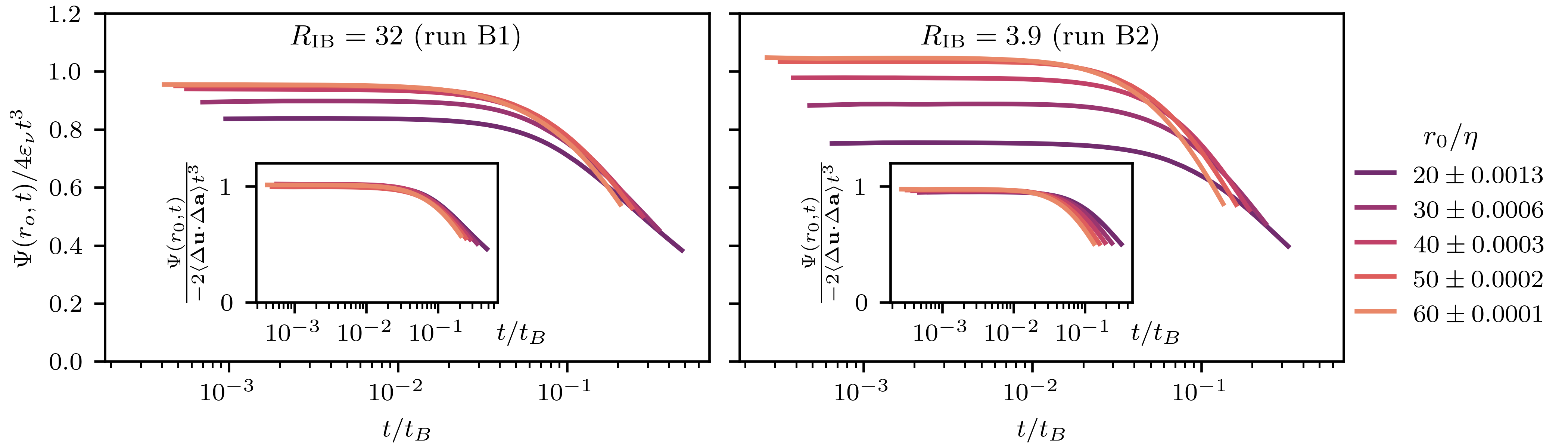}
\caption{Time anti-symmetric part of relative pair dispersion for different initial separations $r_0$ within the inertial range (indicated by color), compensated by the HIT prediction using $\left\langle \Delta_{r_0} \vec{u} \cdot \Delta_{r_0}\vec{a} \right\rangle = - 2 \varepsilon_\nu$. Compare with the previously studied HIT case~\cite{juchaTimereversalsymmetry2014}. The left panel shows run B1, with only small differences to the HIT case~\cite{juchaTimereversalsymmetry2014}, the right panel shows, run B2 in \cref{tab:param} showing a scale dependency of the pair dispersion time asymmetry. The inlets demonstrate the validity of the Taylor expansion in \cref{eq:TaylorExpansionPair} for both cases. Here, we used a different set of pairs to estimate $\Psi(r_0,t)$ and $\left\langle \Delta_{r_0} \vec{u} \cdot \Delta_{r_0}\vec{a} \right\rangle$ to avoid biasing the analysis.}
\label{fig:pairDispersion}
\end{figure*}

While solving the Boussinesq equations (\cref{eq:Bous_u,eq:Bous_T}), we follow the trajectories of $N_p \gtrsim 10^6$  tracer particles, initially randomly distributed throughout the system. To obtain information of particle pairs from trajectories both forwards and backwards in time, we identify, at specific times, pairs of particles whose relative distance is within a chosen range of distance $r - \Delta r \leq |\vec{r}| \leq r + \Delta r$ with the tolerance $\Delta r$. For large databases, the computation time to compute all relative distances grows rapidly. For an efficient parallelized  approach to this challenge we refer to~\cite{buariaCharacteristics2015}. Here, we focus on single processor optimization by employing a novel algorithm using hierarchical spatial domain partitioning with an octree data structure~\cite{finkelQuad1974}. In test runs with synthetic data, we observed a reduction of computational time by a factor of up to $10^3$, depending on various parameters such as $N_p$, $r$ and $\Delta r$. For a short description of the algorithm we refer to the supplemental material~\cite{supp}, for more details see~\cite{gallonPhd}. For the analysis presented in the following, we chose the tolerances $\Delta r_0 \propto {r_0}^{-2}$ to keep the number of pairs independent of $r_0$. As a reference, we chose $\Delta r_0 = 0.005 \eta$ for $r_0 = 10\eta$. We averaged over $\sim 10^6$ pairs per run and distance $r_0$ that are taken, assuming a statistically stationary flow, from $256$ snapshots of the simulation.

A measure of the relative dispersion of two particles is provided by the evolution of the squared separation $\delta_{r_0} r(t)^2 = \qty(\vec{r}(t) - \vec{r}_0)^2$, imposing that the initial separation $\vec{r}(0) = \vec{r}_0$~\cite{batchelorApplication1950,bourgoinRole2006,ouelletteExperimental2006,juchaTimereversalsymmetry2014,polancoMultiparticle2023}. Focusing on short times, a straightforward Taylor series expansion leads to:
\begin{equation}\label{eq:TaylorExpansionPair}
    \delta_{r_0} r(t)^2 \approx \qty(\Delta_{r_0} \vec{u})^2 t^2 + \qty(\Delta_{r_0} \vec{u} \cdot \Delta_{r_0} \vec{a}) t^3 \qq*{,}
\end{equation}    
where $\Delta_{r_0}\vec{u}$ and $\Delta_{r_0}\vec{a}$ denote the initial value of the pairs relative velocity and acceleration respectively~\cite{batchelorApplication1950}. Note that \cref{eq:TaylorExpansionPair} holds for the motion of particle pairs in any Newtonian system. By averaging over all pairs of particles with the same initial separation,  the cubic term in \cref{eq:TaylorExpansionPair} provides a direct connection with the intrinsic time-asymmetry of the flow, as it was the case for HIT~\cite{juchaTimereversalsymmetry2014}. Inserting \cref{eq:TaylorExpansionPair} we obtain, for the function $\Psi$ characterizing the asymmetry of $\delta_{r_0} r(t)^2$: 
\begin{align}
    \Psi(r_0,t) \equiv \langle \delta_{r_0} r^2(-t) - \delta_{r_0} r^2(t)  \rangle 
    \approx -2 \langle \Delta_{r_0} \vec{u} \cdot \Delta_{r_0} \vec{a} \rangle t^3 \text{.}
    \label{eq:pairsAsym}
\end{align}
The approximate expression \cref{eq:TaylorExpansionPair} is estimated to remain valid over the Bachelor time $t_B(r_0) = \abs{\langle{\qty(\Delta_{r_0} \vec{u})}^2 \rangle /  \langle \Delta_{r_0}\vec{u} \cdot  \Delta_{r_0} \vec{a} \rangle  }$~\cite{batchelorApplication1950,bitaneTime2012}.
For HIT, assuming $\langle\qty(\Delta_{r_0} \vec{u})^2 \rangle \propto \varepsilon_\nu^{2/3} r_0^{2/3}$ and $\langle {\Delta_{r_0}\vec{u}} \cdot \Delta_{r_0} \vec{a} \rangle = - 2 \varepsilon_\nu$, $t_B$ reduces to the Kolmogorov time: $t_B(r_0) \propto \varepsilon_\nu^{-1/3}r^{2/3} = \tau_{r_0}$. For ROTSTRAT turbulence, $\langle\qty(\Delta_{r_0} \vec{u})^2 \rangle$ follows a different scaling law with $r_0$~\cite{alexakisCascades2018,davidsonTurbulence2013}. Here, we merely determine $t_B$ as the ratio between $\langle{\qty(\Delta_{r_0} \vec{u})}^2 \rangle $ and $  \left\langle \Delta_{r_0}\vec{u} \cdot  \Delta_{r_0} \vec{a} \right\rangle $, measured from our own simulations. 

\Cref{fig:pairDispersion} shows that contrary to the HIT case, where $\Psi(r_0,t)/t^3 \approx 4 \varepsilon_\nu$ is independent of $r_0$, provided that $r_0$ is in the inertial range and $t \ll t_B$~\cite{juchaTimereversalsymmetry2014}, in the ROTSTRAT case, $\Psi(r_0,t)/t^3$ is a function of $r_0$, which increases with the initial separation $r_0$, the more so as $\mathrm{Fr}$ decreases.

\begin{figure*}[t]
\includegraphics{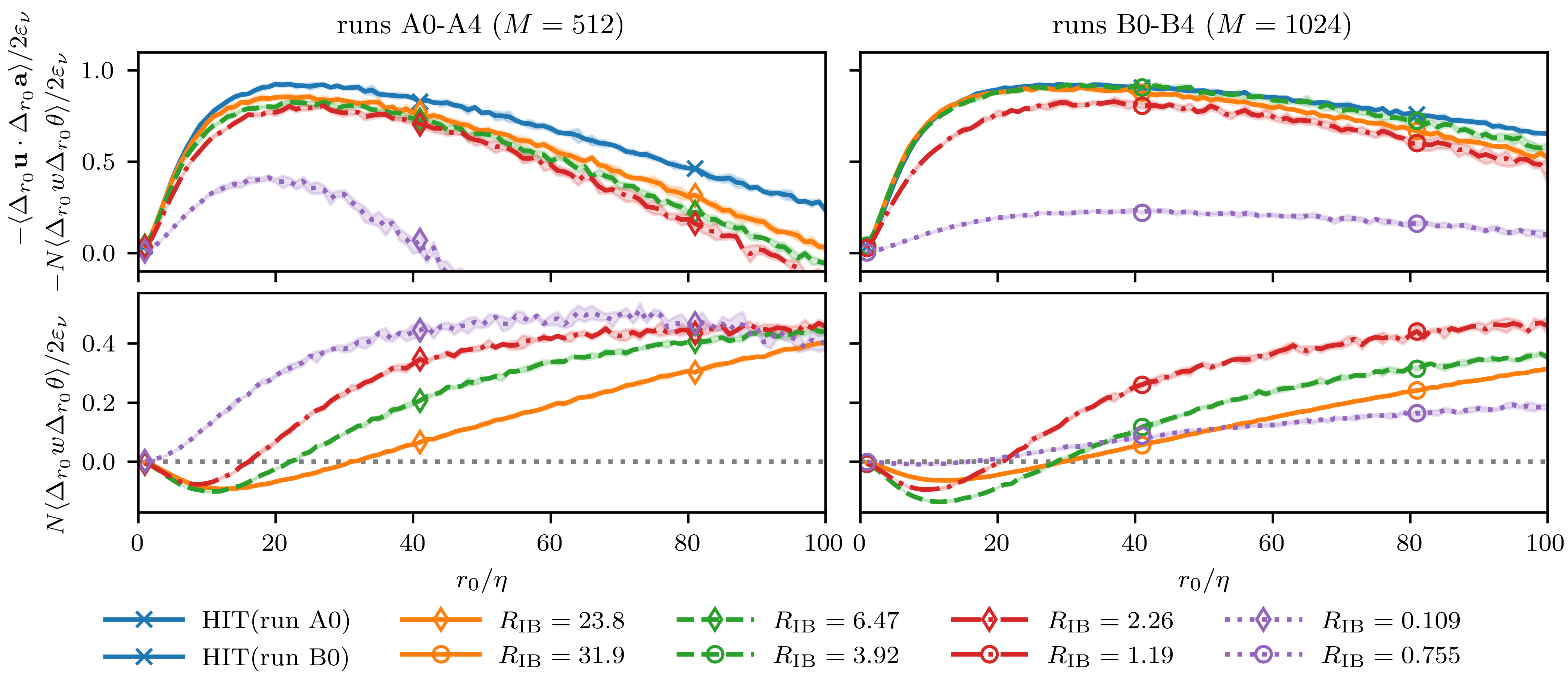}
\caption{Kinetic energy budget (first row) and exchange term (second row) for all runs. The left column shows the observations for the $M=512$ runs (A0-A4) (diamonds), the right column those for the $M=1024$ runs (B0-B4) (circles). Colors indicate different values of $R_\mathrm{IB}$, as indicated in the legend. The shaded areas indicate the standard deviation computed averaging over $4$ subsets. The lines correspond to the average over the whole data set. Furthermore, the dotted gray line in the second row denotes a vanishing energy exchange, i.e. $N \langle \Delta_{r_0}\theta \Delta_{r_0}w\rangle=0$.}
\label{fig:energy-budget}
\end{figure*}

In \cref{eq:pairsAsym}, we identify the average rate of change of the pairs kinetic energy $\langle \mathrm{D}/\mathrm{D}t\ E_\mathrm{kin}^{(r_0)}\rangle = \langle {\Delta_{r_0} \vec{u}} \cdot {\Delta_{r_0} \vec{a}} \rangle$. To proceed, we extend the one-particle energy budget to the pair energy budget. The former is obtained by averaging the Boussinesq equations (\cref{eq:Bous_u,eq:Bous_T}). Assuming homogeneity and stationarity, one finds (see~\cite{supp}):
\begin{align}
\varepsilon_{\Phi} - \varepsilon_\nu - N \langle \theta w \rangle =0 \qq{and}
N \langle \theta w \rangle -\varepsilon_\kappa = 0,
\end{align}
where $\varepsilon_\Phi$ is the energy source, and $\varepsilon_\nu$ and $\varepsilon_\kappa$ are respectively the dissipation terms for the kinetic and potential energy. The exchange term $N \langle \theta w \rangle$ represents the amount of kinetic energy converted into potential energy. This term is positive
and equal to the amount of dissipated potential energy $\varepsilon_\kappa$~\cite{sozzaDimensional2015}. The Lagrangian two-point energy budget, describing the transfer of energy between two particles, generalizes the classical Karman-Howarth-Monin relations. By averaging over the Boussinesq equations (\cref{eq:Bous_u,eq:Bous_T}), assuming homogeneity and stationarity~\cite{mannExperimental1999,hillExact2002,sozzaDimensional2015}, we obtain (see~\cite{supp}): 
\begin{align}
    \left\langle {\Delta_{r_0}\vec{u}} \cdot {\Delta_{r_0} \vec{a}} \right\rangle
    &= - 2 \varepsilon_\nu - N \langle \Delta_{r_0}\theta \Delta_{r_0}w\rangle 
      - 2 D_\nu(r_0) \text{,}
    \label{eq:khm-kin} \\
    \left\langle {\Delta_{r_0}{\theta}} \cdot \Delta_{r_0} \dot\theta \right\rangle
    &= - 2 \varepsilon_\kappa + N \langle \Delta_{r_0}\theta \Delta_{r_0} w\rangle - 2 D_\kappa(r_0)
    \label{eq:kmh-pot} \text{,}
\end{align}
with the mixed dissipation terms $D_\nu(r_0) = \nu \left\langle \sum_{i,j} \pdv{u_i(\vec{x})}{x_j}\pdv{u_i(\vec{x}+\vec{r})}{x_j}\right\rangle$ and  $D_\kappa(r_0) = \kappa \left\langle \sum_{j} \pdv{\theta(\vec{x})}{x_j}\pdv{\theta(\vec{x}+\vec{r})}{x_j}\right\rangle $. We assume the gradients to decorrelate quickly for sufficiently large separations $r_0$, such that the mixed dissipation terms can be neglected. In the absence of stratification, the Karman-Howarth-Monin relation reduces to the well-known relation $\left\langle {\Delta_{r_0}\vec{u}} \cdot {\Delta_{r_0} \vec{a}} \right\rangle = - 2 \varepsilon_\nu$., valid in HIT. Furthermore, we note that all rotational terms vanish.

The signs of the various terms in \cref{eq:khm-kin,eq:kmh-pot} determine the direction of the transfer. The negative sign of $-2\varepsilon_\nu$ is the signature of the presence of a direct energy cascade. We remark that in two-dimensional turbulence, the sign of the cross-correlation between acceleration and velocity is inverted. This is a consequence of the correlation between velocity and dissipation at large scales, whereas forcing acts at a much smaller scale, from which energy flux  arranges in an inverse cascade~\cite{bernardThreepoint1999,braggIrreversibility2018}.

In our simulations, we observe the sum $-  [ \langle \Delta_{r_0} \mathbf{u} \cdot \Delta_{r_0} \mathbf{a} \rangle + N \langle \Delta_{r_0} w \Delta_{r_0} \theta \rangle ] / 2 \varepsilon_\nu$ to be constant and close to unity over a certain range of scales (which increases with the $\mathrm{Re}$) for runs with  $R_\mathrm{IB} > 1$, as shown in the first row of \cref{fig:energy-budget}. According to \cref{eq:khm-kin}, this term is equal to  $1 + D_\nu(r_0)/(2 \varepsilon_\nu)$ and can thus be seen as the full kinetic energy budget. For the cases considered here with $R_\mathrm{IB} <1$, we observe a stronger deviation from $1$, possibly pointing to a transition to a different energy transfer mechanism.

Judging from \cref{eq:khm-kin,eq:kmh-pot}, the term $N\langle \Delta_{r_0}\theta \Delta_{r_0} w\rangle$, represents an average exchange from kinetic to potential energy if positive, or conversely if negative.  In our simulations, we observe for large separations in the inertial range the average energy exchange to be positive, i.e. converting kinetic energy to potential energy. For small separations and $R_\mathrm{IB}>1$, however, we observe a negative average energy exchange, hence transferring energy in the opposite direction, i.e. from potential to kinetic energy, as shown in the second row of \cref{fig:energy-budget}. This result is consistent with other observations from Eulerian studies in spectral space~\cite{hollowayBuoyancy1988,staquetStatistical1998,carnevaleBuoyancy2001,brethouwerScaling2007,roraiStably2015}.

\begin{figure}[t]
\includegraphics{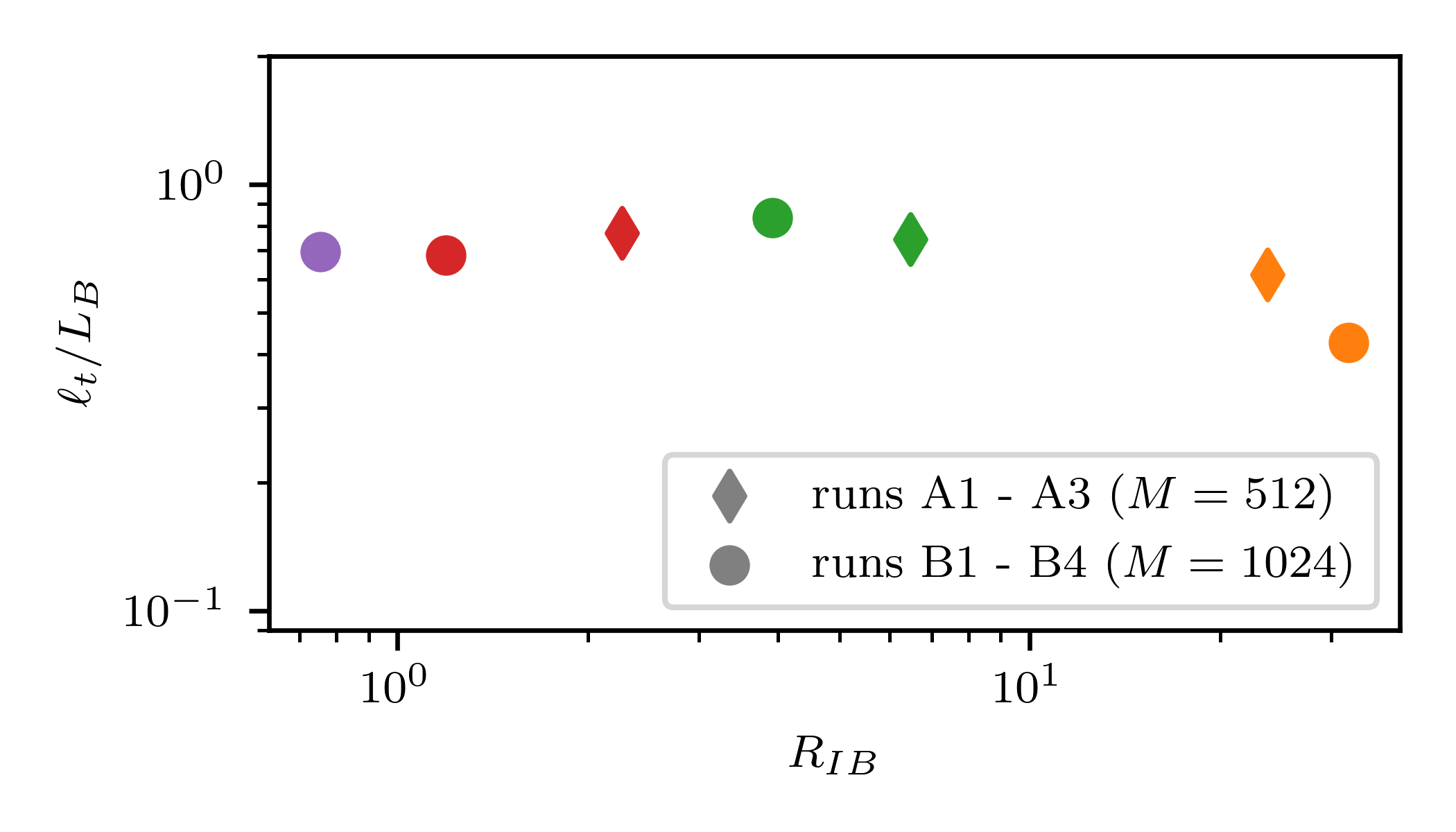}
\caption{Length scale $\ell_t$ at which the exchange term changes compensated by the buoyancy length scale $L_B$ as a function of $R_{\mathrm{IB}}$} \label{fig:zero-crossing}
\end{figure}

We now consider the transition length scale $\ell_t$, where the change of sign occurs, i.e. $N \langle \Delta_{\ell_t} \theta \Delta_{\ell_t} w \rangle=0$. \Cref{fig:zero-crossing} shows the ratio between $\ell_t$ and the buoyancy length-scale $L_B$ as a function of $R_\mathrm{IB}$. Although the ratio $\ell_t/L_B$ is not strictly constant, it remains of order $1$ throughout the simulations with $R_{\mathrm{IB}}>1$, where $0.035 \lesssim \mathrm{Fr} \lesssim 0.17$. This suggests that the mechanism leading to the energy exchange results from the complex interactions between internal gravity waves and turbulent eddies. The connection even prevails very close to the transition at $R_\mathrm{IB}=1$ (see run B3 and B4). At even lower $R_\mathrm{IB}$, we see a transition to a different regime. Furthermore, at the largest value of $R_{\mathrm{IB}}$ studied here, $\ell_t/L_B$ deviates more from $\sim 1$ than for the runs at lower values of $R_{\mathrm{IB}}$. In addition, we remark that run B1 (full, yellow line with circles in \cref{fig:energy-budget}) shows a reduced exchange term at small scales compared to the other runs. This may indicate a different behavior, due to the large value of $R_{\mathrm{IB}}$. 

 The observed energy exchange directly influences the time asymmetry in particle dispersion $\Psi(r_0,t)$, and explains the differences to the previously studied case of HIT~\cite{juchaTimereversalsymmetry2014}. For separations $ r_0 > L_B$, the average energy transfer is directed from kinetic to potential energy and increases with $r_0$. This explains the observed increase of $\Psi(r_0,t)$ with increasing $r_0$ in \cref{fig:pairDispersion}. For separations $ \eta < r_0 < L_B$, the energy transfer is from potential to kinetic energy, symmetrically increasing and decreasing pair dispersion backwards and forwards in time and are therefore reducing the pair asymmetry $\Psi(r_0,t)$.

A precise description of the energy exchanges in rotating stratified turbulent flows is crucial to understand the interplay between waves and eddies. Remarkably, our work stresses the role of the buoyancy length scale, $L_B$, over a broad range of $R_{\mathrm{IB}}$ from an eddy dominated regime ($R_{\mathrm{IB}} \gtrsim 10)$~\cite{Pouquet+19} until the transitional regime to wave-dominated flows ($R_{\mathrm{IB}} \lesssim 1$). Interesting questions for future investigation are whether our conclusions extend to the range $100 \lesssim R_{\mathrm{IB}} \lesssim 1000$, corresponding to measurements in the oceans~\cite{Moum:1996} and to similar values of $R_{\mathrm{IB}}$ at higher $\mathrm{Re}$ and lower $\mathrm{Fr}$. We conclude by noticing that the time-asymmetry of the relative energy between two tracer particles could be used to measure the energy dissipation rate, $\varepsilon_\nu$, in terrestrial or planetary flows~\cite{pearsonAdvective2021,pearson2023}. In this context, understanding of the exchange between kinetic and potential energy in such flows appears as an important step.

\begin{acknowledgments}
S.G. acknowledges the "Fond Recherche" of ENS Lyon for financial support. A.S. acknowledges support from the post-doctoral fellowship program LABEX MILYON of the Universit\'e de Lyon. R.M. and F.F. acknowledge support from the project ``EVENTFUL'' (ANR-20-CE30-0011), funded by the French ``Agence Nationale de la Recherche'' - ANR through the program AAPG-2020.x
The computing resources utilized in this work were provided by PSMN at the \'Ecole Normale Superieure de Lyon and PMCS2I at the \'Ecole Centrale de Lyon.
\end{acknowledgments}

\nocite{landauFluid1987}
\nocite{brethouwerScaling2007}
%

\clearpage

\end{document}


\title{Supplemental Material for: \\
Lagrangian irreversibility and energy exchanges in rotating-stratified turbulent flows}

\author{S. Gallon}
\affiliation{Univ Lyon, ENS de Lyon, CNRS, Laboratoire de Physique, F-69342 Lyon, France}
\author{A. Sozza}
\affiliation{Univ Lyon, ENS de Lyon, CNRS, Laboratoire de Physique, F-69342 Lyon, France}
\affiliation{CNR, Institute of Atmospheric Sciences and Climate, 10133 Torino, Italy}
\author{F. Feraco}
\affiliation{Univ Lyon, CNRS, École Centrale de Lyon, INSA de Lyon, Univ Claude Bernard Lyon 1, Laboratoire de Mécanique des Fluides et d’Acoustique - UMR 5509, F-69134 Écully, France}
\affiliation{Leibniz-Institute of Atmospheric Physics at the Rostock University, 18225 Kühlungsborn, Germany}
\author{R. Marino}
\affiliation{Univ Lyon, CNRS, École Centrale de Lyon, INSA de Lyon, Univ Claude Bernard Lyon 1, Laboratoire de Mécanique des Fluides et d’Acoustique - UMR 5509, F-69134 Écully, France}
\author{A. Pumir}
\affiliation{Univ Lyon, ENS de Lyon, CNRS, Laboratoire de Physique, F-69342 Lyon, France}
\affiliation{Max Planck Institute for Dynamics and Self-Organization, 37077 Göttingen, Germany}

\date{January 25, 2024}
\maketitle
\section{Octree based algorithm for particle pair identification}
Experimental and numerical studies of turbulent flows often generate very large databases. The present investigation in Lagrangian turbulence relies on the analysis of millions of particle trajectories. Extracting meaningful information on the structure of the flow, as a function of scale $r$, rests on the capability to identify sets of particles separated by a distance $r$, up to a tolerance $\Delta r$. The corresponding search through the database, however, can be a very expensive task, the more so as the total number of particles $N_p$, increases. This calls for the development of efficient methods to search through the database.

A way to reduce computational costs is to divide the domain into sub-domains, e.g. into sub-cubes of edge length $\ell$ and sort all the particles into these sub-cubes. When $r + \Delta r < \ell$, the particles within a distance between $r - \Delta r$ and $r + \Delta r$ from a given particle $p_1$ are either in the same sub-cube as $p_1$ or in one of the neighboring $26$ sub-cubes. Restricting the search to these sub-cubes can lead to a significant reduction of the computational effort. To further improve the performance especially for larger separations $r$ between particles, we propose a search algorithm based on the octree data structure \cite{finkelQuad1974}. Within this approach, the original cubic domain, $[0,2\pi]^3$, is subdivided into 8 sub-cubes with each half the original side length. The original cube is referred to as the root node and the sub-cubes as its children. Each of the children nodes is then itself recursively subdivided into 8 sub-cubes per generation, with each half the side length of the previous generation. The process is repeated for a total of $m$ times. The particles are then sorted into the $8^m$ smallest sub-cubes (leaf nodes). For the present application, we assume the particles to be uniformly distributed (note that for some other applications of the octree-data structure, the number of refinements is chosen locally based on the local particle density). To choose the number of refinements $m$, we found that the method works efficiently with at least $\sim 10$ particles per cube on the smallest level. 

\begin{figure}
    \centering
\includegraphics{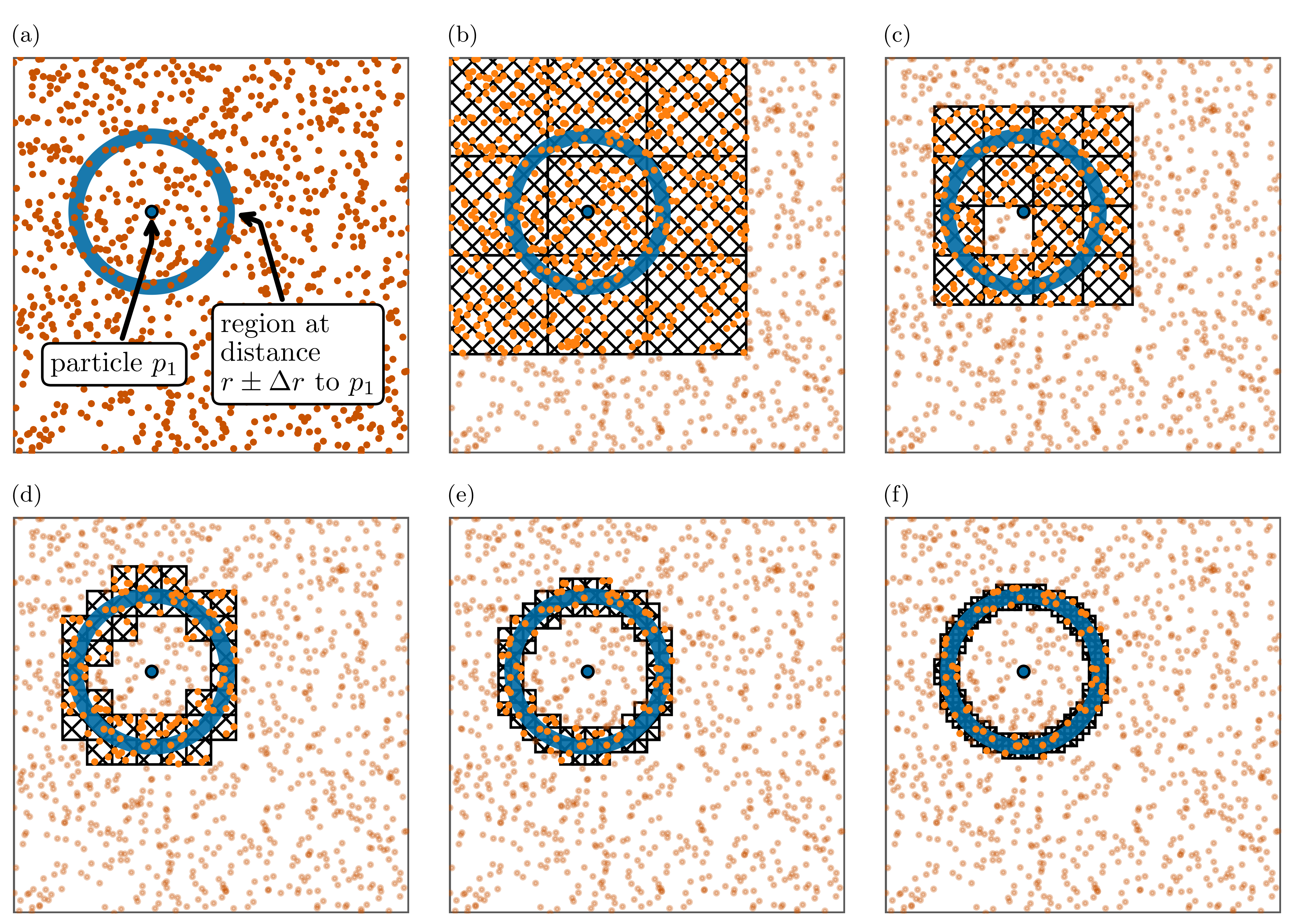}
    \caption{Exemplary depiction of the pair search algorithm using a quadtree-data structure in two dimensions. (a-f) The blue point with black edge is the particle $p1$, the shell of distances $r \pm \Delta r$ is colored in blue. (b-f) Iterative search using the quadtree data structure. 
    The hatched regions are the nodes that intersect with the shell in the respective generation. 
    }
    \label{fig:quadtreePairSearch}
\end{figure}

Particles at a separation within $[ r - \Delta r, r + \Delta r]$ to a particle $p_1$ are located in a shell, as represented for the two-dimensional case in \cref{fig:quadtreePairSearch}(a). To find these particles, we start at the root node and descent in the tree to the child node that contains the position of $p_1$. We repeat this descent until we arrive at a node whose side length $\ell_i$ fulfills $\ell_i/2 < r + \Delta r < \ell_i$, which is the node with the smallest side length that is still larger than the maximum search distance $r + \Delta r$. All particles that are at distance $r \pm \Delta r$ to $p_1$ can therefore be only either within this node, or within one of its neighbors , see \cref{fig:quadtreePairSearch}(b). We then look for all the nodes of smaller size that intersect with the shell by the following recursive approach. A given node can contain a particle at a distance between $r-\Delta r$ and $r + \Delta r$ from a particle $p_1$ only if it has an intersection with the shell represented in \cref{fig:quadtreePairSearch}. Therefore, if a node is either too close or too far away from $p_1$, itself and its children can be discarded. The children of the nodes that have an intersection with the shell are considered in the next step, using the same procedure until the leaf nodes are reached (see \cref{fig:quadtreePairSearch}(c-f)). To finalize the search, we compute the distances of the particles in the remaining nodes to $p_1$ to find all pair partners. This procedure is repeated for all $p_1$ to find all possible pairs. To avoid double counting, we only keep pairs with increasing indices, i.e. $p_1 < p_2$.
\section{Derivation of the Pair Energy Budget}
In this work, we consider a stably stratified fluid with a linear density profile $\overline{\rho}=\rho_0 -\gamma z$ 
in a rotating frame of reference with solid body rotation 
rate $\Omega$ (and frequency $f=2\Omega$).
In this case, the equations of motion for the incompressible velocity field $\vec{u}=(u,v,w)$, i.e. $\div \uu = 0$ and density fluctuation $\theta=(\gamma/N)(\rho-\overline{\rho})$ read
\begin{align}
\label{eq:Bous_u}
\partial_t \vec{u}  + \vec{u} \cdot \boldsymbol{\nabla} \vec{u} &=  -\grad p - f \vec{e}_z  \cross \vec{u} - N \theta \vec{e}_z + \nu \laplacian \vec{u} + \vec{\Phi} \qq*{,} \\
\partial_t \theta + \vec{u} \cdot \grad \theta &= 
N w + \kappa \laplacian \theta \qq*{,}\label{eq:Bous_T} 
\end{align} with the reduced pressure $p$,
the Brunt-V\"ais\"al\"a frequency 
$N=(\gamma g/\rho_0)^{1/2}$, the kinematic viscosity $\nu$, the external forcing $\Phi$ and the thermal diffusivity $\kappa$. 

In the following, we derive the energy budget for the relative kinetic and potential energy for two particles separated by $\vec{r}$. For the original Eulerian work, we refer to \cite{sozzaDimensional2015}.

\subsection{One Particle Energy Budget}

We start by considering a single particle with velocity $\vec{u}$ and acceleration $\vec{a} = \advDeriv \uu = \partial_t \vec{u} + \vec{u} \cdot \boldsymbol{\nabla} \vec{u}$. The rate of change of the kinetic energy $E_\mathrm{kin} =\vec{u}^2/2$ of this particle along its trajectory is given by
\begin{equation}
  \advDeriv E_\mathrm{kin} = \vec{u} \cdot \vec{a} \qq*{.}
\end{equation}
Averaging this quantity over many particles and inserting the equation of motion, we obtain
\begin{align}
\average{\advDeriv E_\mathrm{kin}} = - \average{\vec{u} \cdot \grad p} - \average{\vec{u} \cdot f \vec{e}_z  \cross \vec{u}} - \average{\vec{u} \cdot N \theta \vec{e}_z} +  \average{\vec{u} \cdot \nu \laplacian \vec{u}} + \average{\vec{u} \cdot \vec{\Phi}} \qq*{.} \label{eq:ua}
\end{align}
The first term on the right hand side of \cref{eq:ua} can be rewritten as
\begin{equation}
\average{\vec{u} \cdot \grad p} = \divergence \average{\vec{u} \cdot p} - \average{p \boldsymbol{\nabla} \cdot \vec{u}} = 0 \qq*{,}
\end{equation}
and therefore vanishes due to homogeneity (first term) and incompressibility (second term). 

The second term on the right hand side of \cref{eq:ua} is zero for geometric reasons ($\vec{u} \cdot \vec{e}_z  \cross \vec{u} = 0$).

The third term on the right hand side of \cref{eq:ua} represents the energy exchange between potential and kinetic energy. For a more compact notation, we introduce 
\begin{equation}
\varepsilon_\chi \equiv \average{\vec{u} \cdot N \theta \vec{e}_z} = N \average{w \theta} \qq*{,}
\end{equation}
where $w$ denotes the vertical velocity component. With our sign conventions, $\varepsilon_\chi$ represents a sink of kinetic energy and a source of potential energy for $\varepsilon_\chi > 0$. 

The fourth term on the right hand side of \cref{eq:ua} describes the energy dissipation by viscosity
\begin{align}
\average{\vec{u} \cdot \nu \laplacian \vec{u}} &= \frac{\nu}{2} \laplacian \average{ \qty(\vec{u})^2 } - \varepsilon_\nu  \qq*{.}
\end{align}
The first term on the right hand side of the equation above vanishes due to homogeneity. Furthermore we introduced the kinetic energy dissipation rate
\begin{equation}
\varepsilon_\nu \equiv \nu \average{\sum_{i,j} \qty(\pdv{u_i}{x_j})^2} \qq*{.}    
\end{equation}

Finally, the remaining term on the right hand side of \cref{eq:ua} is the insertion of energy due to the large scale forcing $\vec{\Phi}$. For a more compact notation, we introduce
\begin{equation}
  \varepsilon_\Phi \equiv \average{\vec{u} \cdot \vec{\Phi}}  \qq*{.}
\end{equation}
Putting everything together, we arrive at the one point budget for the kinetic energy
\begin{equation}
    \average{\advDeriv E_\mathrm{kin}} = \langle \uu \cdot \aa \rangle = \varepsilon_\Phi - \varepsilon_\chi - \varepsilon_\nu  \qq*{.} \label{eq:1PBudgetKin}
\end{equation}
For the potential energy $E_\mathrm{pot} = \theta^2/2$, we rewrite the thermal diffusive term similarly to the viscous term and directly obtain similarly the budget 
\begin{equation}
    \average{\advDeriv E_\mathrm{pot}} = \langle \theta \dot\theta \rangle =  \varepsilon_\chi - \varepsilon_\kappa  \qq*{,} \label{eq:1PBudgetPot}
\end{equation}
with the potential energy dissipation rate 
\begin{equation}
\varepsilon_\kappa \equiv \kappa \average{\sum_{j} \qty(\pdv{\theta}{x_j})^2} \qq*{.}
\end{equation}

\subsection{Two Particle Energy Budget}

We now consider a pair of particles with positions $\vec{x}^{(1)}(t)$ and $\vec{x}^{(2)}(t)$, separated by a displacement $\vec{r}(t)$, where $\abs{\vec{r}(0)}=r$ is in the inertial range, and study the evolution of the relative kinetic energy of this pair $E^{(\vec{r})}_\mathrm{kin}= \Delta_\vec{r} \uu^2 / 2$: 
\begin{equation}
\advDeriv E_\mathrm{kin}^{(2)} = \Delta_\vec{r} \vec{u} \cdot \Delta_\vec{r} \vec{a} \qq*{,}
\label{eq:dE2_dt}
\end{equation}
where we have used the notation $\Delta_\vec{r} q \equiv q^{(2)} - q^{(1)}$ for any observable $q$. Averaging \cref{eq:dE2_dt} over many pairs with the \textit{same} separation $\vec{r}$, and using the Boussinesq equations, one obtains:
\begin{equation}
\average{\advDeriv E_\mathrm{kin}^{(2)}}_\vec{r}
= \average{\vec{u}^{(2)} \cdot \vec{a}^{(2)}}_\vec{r} - \average{\vec{u}^{(2)} \cdot \vec{a}^{(1)}}_\vec{r} - \average{\vec{u}^{(1)} \cdot \vec{a}^{(2)}}_\vec{r} + \average{\vec{u}^{(1)} \cdot \vec{a}^{(1)}}_\vec{r} ~ .
\label{eq:duda}
\end{equation}
Note that we are not only prescribing here the length of $\vec{r}$, but also its orientation. 
The first and last term on the right hand side of \cref{eq:duda} are the rates of change of the kinetic energy of each of the two particles separately. Because of statistical homogeneity, they are independent of the initial separation $\vec{r}$ and therefore each term is given by the single particle budget for the kinetic energy derived above:
\begin{equation}
\average{\vec{u}^{(2)} \cdot \vec{a}^{(2)}}_\vec{r} = \average{\vec{u}^{(1)} \cdot \vec{a}^{(1)}}_\vec{r} = \varepsilon_\Phi - \varepsilon_\chi - \varepsilon_\nu \qq*{.}
\end{equation}
To evaluate the mixed terms on the right hand side of \cref{eq:duda}, we first use arguments similar to those used for the one-particle energy budget to show that pressure terms do not contribute. As an example:
\begin{align}
\average{\vec{u}^{(2)} \grad^{(1)} p^{(1)}}_\vec{r} = \boldsymbol{\nabla}^{(1)}\cdot \average{\vec{u}^{(2)} p^{(1)}}_\vec{r} - \average{p^{(2)} \boldsymbol{\nabla}^{(2)} \cdot \vec{u}^{(1)}}_\vec{r} = 0 \qq*{,}
\end{align}
where, we used $\boldsymbol{\nabla}^{(2)} \cdot \vec{u}^{(1)} = \boldsymbol{\nabla}^{(1)} \cdot \vec{u}^{(1)} = 0$. 

The sum of the mixed terms involving the rotational (Coriolis force) vanishes as a consequence of the anti-symmetry of the cross product
\begin{align}
\average{\vec{u}^{(2)} \cdot f \vec{e}_z  \cross \vec{u}^{(1)}}_\vec{r} + \average{\vec{u}^{(1)}\cdot f \vec{e}_z  \cross \vec{u}^{(2)}} = f \vec{e}_z \cdot \average{\vec{u}^{(2)} \cross \vec{u}^{(1)} + \vec{u}^{(1)}\cross\vec{u}^{(2)}} =  0 \qq*{.}
\end{align}
The mixed terms involving the density fluctuation $\theta$ cannot be eliminated. Together with the contribution from the single particle budgets of each of the particles, we obtain
\begin{align}
2\varepsilon_\chi - \average{w^{(2)}  N \theta^{(1)}}_\vec{r} - \average{w^{(1)}  N \theta^{(2)} }_\vec{r} &= N \average{\Delta_\vec{r}w \Delta_\vec{r} \theta}_\vec{r} \qq*{,}
\end{align}
In this work, we consider a forcing on the large scales. We thus assume it to vary little for sufficiently small $r$ and write $\vec{\Phi}^{(2)} \approx \vec{\Phi}^{(1)}$. The forcing contribution from the mixed particle terms thus read
\begin{equation}
     - \langle \uu^{(2)} \cdot \vec{\Phi}^{(1)} \rangle - \langle \uu^{(1)} \cdot \vec{\Phi}^{(2)} \rangle \approx - 2\varepsilon_\Phi
\end{equation}
and cancel out the $+2\varepsilon_\Phi$ from the single particle terms.

To simplify the mixed viscous terms, we rewrite their sum 
\begin{equation}
\average{\vec{u}^{(2)} \cdot \nu \laplacianWithIndex{(1)}
\vec{u}^{(1)}}_\vec{r} +  \average{\vec{u}^{(1)} \cdot \nu \laplacianWithIndex{(2)}\vec{u}^{(2)}}_\vec{r}
= \nu \laplacianWithIndex{(1)} \average{\vec{u}^{(1)}\vec{u}^{(2)}}_\vec{r}  - 2 D_\nu(\vec{r}) \qq*{,} \label{eq:PairViscous}
\end{equation}
where the first term on the right hand side vanishes due to homogeneity and with the mixed dissipation term
\begin{equation}
    D_\nu(\vec{r}) \equiv \nu \average{\sum_{i,j} \qty(\pdv{u_i^{(1)}}{x_j^{(1)}})\qty(\pdv{u_i^{(2)}}{x_j^{(2)}})}_\vec{r} \qq*{.}
\end{equation}

Putting everything together, we obtain the two point budget for the kinetic pair energy
\begin{align}
\average{\advDeriv E_\mathrm{kin}^{(2)}}_\vec{r} &=  \langle \Delta_\vec{r}\uu \cdot \Delta_\vec{r} \aa \rangle_\vec{r} 
= - 2\varepsilon_\nu - N \average{\Delta_\vec{r}w \Delta_\vec{r} \theta}_\vec{r}  -  2D_\nu(\vec{r})
\qq*{.} \label{eq:2PBudgetKin}
\end{align}
Similarly, we obtain the two particle budget for the potential pair energy $E_\mathrm{pot}^{(2)} = \theta^2/2$:
\begin{align}
 \average{\advDeriv E_\mathrm{pot}^{(2)}}_\vec{r} =     \left\langle {\Delta_{\vec{r}}{\theta}} \cdot \Delta_\vec{r} \dot\theta \right\rangle
    = - 2 \varepsilon_\kappa + N \langle \Delta_\vec{r}\theta \Delta_\vec{r} w\rangle  - 2 D_\kappa(\vec{r}) \qq*{.} \label{eq:2PBudgetPot}
\end{align}
with the mixed thermal diffusion term 
\begin{equation}
    D_\kappa(\vec{r}) \equiv \kappa \left\langle \sum_{j} \pdv{\theta^{(1)}}{x_j}\pdv{\theta^{(2)}}{x_j}\right\rangle_\vec{r} \qq*{.}
\end{equation}

\subsection{Comment on Isotropy}

We note that in the derivation above, we did not explicitly assume the system to be isotropic. In fact, the flows we are considering here are largely anisotropic. In classical textbooks for the case of Navier-Stokes turbulence, the velocity-pressure gradient correlations are normally eliminated under the assumption of isotropy~\cite{frischTurbulence1995,landauFluid1987}. It was then later demonstrated, that the derivation can alternatively be performed by assuming solely homogeneity and incompressibility~\cite{mannExperimental1999,hillExact2002,hillOpportunities2006}.
Furthermore, we note that in the derivation of the pair energy budget, we explicitly fixed the orientation of the separation vector $\vec{r}$. The energy budgets are thus valid when averaging for any fixed distance and orientation, e.g. by prescribing the angle $\vartheta$ between the two particles and the $z$-axis, or when averaging over all orientations as we do in the present work.

\subsection{Comment on Stationarity}

In the derivation above, we did not explicitly assume stationarity. \Cref{eq:1PBudgetKin,eq:1PBudgetPot,eq:2PBudgetKin,eq:2PBudgetPot} should thus also hold at any point in time in a non-stationary system, such as decaying turbulence. Note that then, quantities on both sides of the equations are time-depended.

In our simulations, we have considered statistically steady flows. For this reason, we expect that all quantities on the right hand side of \cref{eq:1PBudgetKin,eq:1PBudgetPot,eq:2PBudgetKin,eq:2PBudgetPot} reach a statistical equilibrium.
Assuming ergodicity, we compute the Lagrangian averages over particle positions and time, as pointed out in the main manuscript. 

In a stationary state, the one particle energy budgets simplify to
\begin{align}
    \average{\advDeriv E_\mathrm{kin}} &= 0 = \langle \uu \cdot \aa \rangle = \varepsilon_\Phi - \varepsilon_\chi - \varepsilon_\nu  \qq*{,} \\
    \average{\advDeriv E_\mathrm{pot}} &= 0 = \langle \theta \dot\theta \rangle =  \varepsilon_\chi - \varepsilon_\kappa \qq*{.}
\end{align}
Hence, velocity and acceleration are correlated in such a way that their scalar product on average is zero.

The left-hand side of the two-particle energy, using the Lagrangian point of view, remains non-zero, even when the flow is statistically steady, as the cross-correlation between velocity and acceleration is not vanishing. We stress that in this case, the identity 
\begin{equation}
    \left\langle \Delta_\vec{r}\uu \cdot \Delta_\vec{r} \aa \right\rangle_\vec{r} = \frac{1}{2} 
\vec{\nabla}\cdot \left\langle |\Delta_\vec{r}\uu|^2\Delta_\vec{r}\uu \right\rangle_\vec{r}
\end{equation}
establishes a connection with the Karman-Howarth-Monin in the Eulerian framework~\cite{sozzaDimensional2015}. 

%